\documentclass[twocolumn,tighten]{aastex63}
\usepackage{amssymb, amsmath,framed}

\usepackage{bm}
\expandafter\ifx\csname package@font\endcsname\relax\else
 \expandafter\expandafter
 \expandafter\usepackage
 \expandafter\expandafter
 \expandafter{\csname package@font\endcsname}
\fi
\def\bq{\begin{equation}}
\def\eq{\end{equation}}
\def\bqy{\begin{eqnarray}}
\def\eqy{\end{eqnarray}}

\def\p{\partial}

\def\p{\partial}

\def\R{\mathbb{R}}

\def\calo{\mathcal{O}}

 \def\q#1#2{q^{\hspace{1pt}#1}_{\,,\hspace{1pt}#2}}
 \def\a#1#2{a^{\hspace{1pt}#1}_{\,,\hspace{1pt}#2}}
 \def\d#1#2{\delta^{\hspace{1pt}#1}_{\,,\hspace{1pt}#2}}

\begin{document}
\title{\large{Atmospheric Escape From TOI-700 d: Venus versus Earth Analogs}}

\correspondingauthor{Chuanfei Dong}
\email{dcfy@princeton.edu}

\author{Chuanfei Dong}
\affiliation{Department of Astrophysical Sciences, Princeton University, Princeton, NJ 08544, USA}
\affiliation{Princeton Plasma Physics Laboratory, Princeton University, Princeton, NJ 08544, USA}

\author{Meng Jin}
\affiliation{SETI Institute, Mountain View, CA 94043, USA}
\affiliation{Lockheed Martin Solar and Astrophysics Lab (LMSAL), Palo Alto, CA 94304, USA}

\author{Manasvi Lingam}
\affiliation{Department of Aerospace, Physics and Space Sciences, Florida Institute of Technology, Melbourne, FL 32901, USA}
\affiliation{Institute for Theory and Computation, Harvard University, Cambridge, MA 02138, USA}

\begin{abstract}
The recent discovery of an Earth-sized planet (TOI-700 d) in the habitable zone of an early-type M-dwarf by the Transiting Exoplanet Survey Satellite constitutes an important advance. In this Letter, we assess the feasibility of this planet to retain an atmosphere -- one of the chief ingredients for surface habitability -- over long timescales by employing state-of-the-art magnetohydrodynamic models to simulate the stellar wind and the associated rates of atmospheric escape. We take two major factors into consideration, namely, the planetary atmospheric composition and magnetic field. In all cases, we determine that the atmospheric ion escape rates are potentially a few orders of magnitude higher than the inner Solar system planets, but TOI-700 d is nevertheless capable of retaining a $1$ bar atmosphere over gigayear timescales for certain regions of the parameter space. The simulations show that the unmagnetized TOI-700 d with a 1 bar Earth-like atmosphere could be stripped away rather quickly ($<$ 1 gigayear), while the unmagnetized TOI-700 d with a 1 bar CO$_2$-dominated atmosphere could persist for many billions of years; we find that the magnetized Earth-like case falls in between these two scenarios. We also discuss the prospects for detecting radio emission of the planet (thereby constraining its magnetic field) and discerning the presence of an atmosphere. \\
\end{abstract}

\section{Introduction} \label{SecIntro}
One of the primary objectives behind detecting and characterizing exoplanets is to seek potentially habitable worlds, and search for biosignatures \citep[e.g.,][]{Kal17,Fuj18}. The first step in this endeavor conventionally entails identifying planets in the so-called ``habitable zone'' (HZ), i.e., the region around the star where a rocky planet can theoretically host liquid water on its surface \citep[e.g.,][]{KWR93,KRK13}. In recent times, many planets have been found in the HZ, with notable examples including Proxima b \citep{AA16} and some planets of the TRAPPIST-1 system \citep{GT17}. On the other hand, it is vital to appreciate that habitability is a multi-faceted concept, and that many factors have to be taken into consideration \citep[e.g.][]{CBB16}. 

One of the most prominent among them is the presence of an atmosphere, which is not only necessary to maintain liquid water on the surface, but also to protect putative biota from high-energy particles and radiation. To this end, non-thermal atmospheric escape mechanisms - many of which are driven by stellar phenomena - have attracted increasing attention \citep[e.g.][]{lammer09,LL19,ABC19}. In consequence, several recent numerical studies based on sophisticated multi-species plasma models have been utilized to estimate the rates of non-thermal atmospheric escape, regulated by stellar winds and space weather events, from water worlds, Earth- and Venus-like planets \citep{GG17,DHL17,DLMC,DJL18,DHL19,EJB19}. 

However, an important point should be appreciated concerning these studies. They have either focused on late-type M-dwarfs (with masses $M_\star \sim$ 0.1 M$_\odot$) or solar-type stars ($M_\star \sim$ 1 M$_\odot$). Yet, a sizable fraction of all stars in the Milky Way presumably fall outside this range \citep[e.g.][]{Kro01}. Hence, it is necessary for studies of atmospheric escape to also encompass early-type M-dwarfs and K-dwarfs. The very recently discovered TOI-700 system seemingly fulfills this criterion admirably, as it comprises a roughly Earth-sized planet (TOI-700 d) in the HZ of an M2 star, whose mass is $M_\star \sim$ 0.42  M$_\odot$ \citep{GBS20,RVZ}. The fact that it was the first Earth-sized planet in the HZ detected by the Transiting Exoplanet Survey Satellite (TESS) accords it further significance.\footnote{\url{https://www.nasa.gov/tess-transiting-exoplanet-survey-satellite}} Lastly, climate models indicate that TOI-700 d may host temperate climates across a diverse array of atmospheric compositions \citep{SWK20}.

In this Letter, we simulate the rates of atmospheric escape from TOI-700 d by tackling a wide palette of configurations. We describe our methodology in Section \ref{SecMeth}, present the ensuing results in Section \ref{SecRes}, and summarize our salient findings in Section \ref{SecConc}.

\section{Methodology}\label{SecMeth}
We describe the state-of-the-art numerical models for calculating stellar wind parameters and rates of atmospheric escape.

\subsection{Stellar wind}\label{SSecSW}
We simulate the stellar wind of TOI-700 using the Alfv\'{e}n Wave Solar Model (AWSoM) within the Space Weather Modeling Framework \citep{toth12}, a data-driven global MHD model initially developed for simulating the solar atmosphere and solar wind \citep{bart14}. AWSoM has been subjected to extensive validation and has accurately reproduced both the solar corona environment (e.g., \citealt{bart14}) and dynamics during coronal mass ejection (CME) eruptions (e.g., \citealt{jin16, jin17}). The model was eventually adapted to simulate stellar winds of several stars (e.g., \citealt{cohen14,DJL18}). 

To adapt the AWSoM for modeling the TOI-700 stellar wind, we utilize the rotational period (54 days), radius (0.420 R$_{\odot}$), and mass (0.416 M$_{\odot}$) of the star from the estimates determined via TESS \citep{GBS20}. Due to the lack of direct surface magnetic-field observations for TOI-700, we employ a solar magnetogram under the solar minimum condition, viz., Global Oscillation Network Group magnetogram of Carrington Rotation 2077 \citep{jin12}; the choice of magnetogram may be justified by the relatively low levels of stellar activity associated with TOI-700 \citep{GBS20}. We scale the mean magnetic flux density to $36$ G, which is estimated based on the X-ray luminosity of the star ($\sim 2.4\times10^{27}$ erg as per \citealt{GBS20}) as well as the solar X-ray luminosity \citep{judge03}. 

\begin{figure*}
\begin{center}
\includegraphics[scale=0.6]{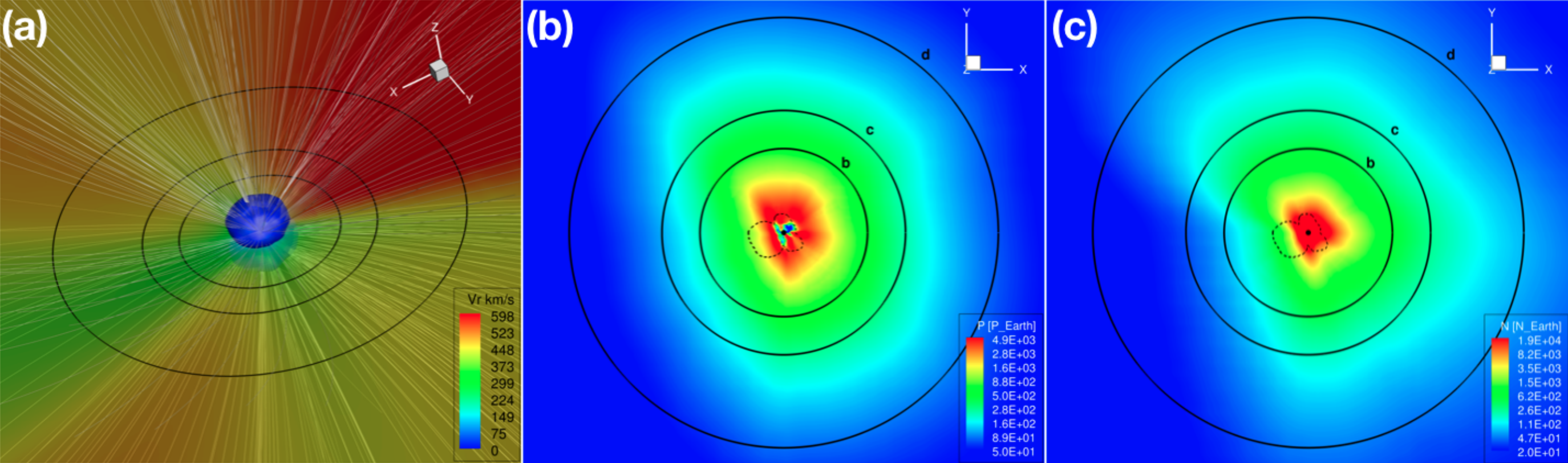}
\end{center}
\caption{Steady-state stellar wind characteristics of TOI-700. (a) The 3D stellar wind configuration comprising select magnetic field lines. Contours in the background illustrate stellar wind speed at the equatorial plane ($z = 0$). The blue isosurface signifies the critical surface beyond which the stellar wind becomes super-magnetosonic (or ``superfast''). Black solid lines indicate the orbits of three planets, namely, TOI-700 b, TOI-700 c, and TOI-700 d. (b) Stellar wind dynamic pressure in the equatorial plane normalized by the solar wind dynamic pressure at 1 AU. (c) Stellar wind density in the equatorial plane normalized by the solar wind density at 1 AU. In panels (b) and (c), the dashed line represents the critical surface location.}
\label{fig:stellarwind}
\end{figure*}

The steady state stellar wind solution is shown in Figure \ref{fig:stellarwind}. The stellar wind speeds at the orbits of TOI-700 planets are comparable to the solar wind speed at 1 AU ($\sim 200$-$600$ km/s). However, because of the much closer distances to the star, the stellar wind density and dynamic pressure are higher, as seen from panels (b) and (c) of Figure \ref{fig:stellarwind}. The critical surface, defined as the region where stellar wind speed is equal to the fast magnetosonic speed, is also computed and shown in Figure \ref{fig:stellarwind}. As with our solar system, all the planets in the TOI-700 system are outside this critical surface, i.e., the stellar wind environment of the planets is always ``superfast''.

\subsection{Atmospheric escape}

To simulate the stellar wind interaction with TOI-700 d, and study resultant atmospheric ion losses, we employ the BATS-R-US multifluid magnetohydrodynamic (MHD) model that was originally developed for solar system planets such as Earth, Venus and Mars \citep{toth12,dong14,dong15,dong18}. The MHD model solves separate continuity, momentum and pressure equations for each ion species. In addition, the electron pressure equation is also evolved separately in our model calculations. The whole upper atmosphere is included in the simulation domain, implying therefore that all types of elastic and inelastic collisions and the associated heating and cooling processes are explicitly present in the MHD equations as source and sink terms. 

For the Venus-like case, we suppose that the 1 bar atmosphere of TOI-700 d is dominated by CO$_2$ analogous to modern Venus. For the Earth-like case, the 1 bar atmosphere of TOI-700 d is akin to that of the modern Earth, i.e., primarily N$_2$ and O$_2$. In this study, we utilize and rescale the 1-D averaged upper atmospheric profiles from the thermosphere general circulation models (GCM) of Venus and Earth \citep{bougher08}. In the case of Venus/Venus-like planets, the dominant neutral species transitions from molecular CO$_2$ to atomic O with increasing altitude in the thermosphere. On the other hand, the entire thermosphere is dominated by atomic oxygen for Earth/Earth-like planets \citep{MWD18}.

It is noteworthy that in addition to photoelectron heating, \emph{Joule heating} (or \emph{Ohmic heating}) is also incorporated in our calculations due to the presence of electron-ion ($\nu_{ei}$) and electron-neutral ($\nu_{en}$) elastic collisions within the electron pressure equation; see \citet{toth12} for more details. Photochemical reactions such as charge exchange, electron impact ionization, and electron recombination inherently comprise a part of the inelastic collision terms, based on the rate coefficients delineated in \citet{schunk2009}. In the case of photoionization, we calculate the rescaled ionization frequencies at TOI-700 d, based on the X-ray luminosity from \citet{GBS20} and the empirical extreme ultraviolet (EUV) scaling from \citet{SMR11}. Note that the optical depth of the upper atmosphere is determined by employing the numerical formula in \citet{smith1972}.

By utilizing a five-species multifluid MHD model, we solve e$^-$, H$^+$, O$^+$, O$_2^+$, CO$_2^+$ for the Venus-like case, and the dominant three fluid species (e$^-$, H$^+$, O$^+$) for the Earth-like case, with these species specified by the dominant ionospheric components \citep{bougher08}; we deliberately adopt a Venus-like atmospheric composition given its importance in exoplanetary science \citep{KAC19}. At the model inner boundary, the species densities satisfy the photochemical equilibrium conditions \citep{schunk2009}. Absorbing boundary conditions are applied to both the velocity and magnetic fields. Due to collisions, both ions and electrons have roughly the same temperature as the neutrals at the inner boundary.

We employ a nonuniform spherical grid for modeling all cases. The angular (i.e., horizontal) resolution is chosen to be 3$^{\circ}$ $\times$ 3$^{\circ}$. However, as the upper atmosphere of a Venus-like planet is cooler than that of a Earth-like planet due to the efficient CO$_2$ cooling \citep{bougher08}, the scale height associated with the former's atmosphere is correspondingly smaller than the latter; therefore, we adopt the smallest radial resolution of 5 km and 10 km for Venus and Earth analogs, respectively, to span the whole upper atmospheric region. The code is run in the Planet-Star-Orbital (PSO) coordinate system, where the $x$ axis is directed from the planet toward the star, the $z$ axis is perpendicular to the planet's orbital plane, and the $y$ axis completes the right-hand system. The simulation domain is set by $-90 \leq x/R_p \leq 45$, $-90 \leq y/R_p \leq 90$ and $-90 \leq z/R_p \leq 90$, where $R_p$ denotes the planetary radius.

We also analyze the role of the planetary magnetic field for the Earth-like case by turning the global dipole magnetic field on and off; to facilitate direct comparison, the dipole moment is taken to be the same as modern Earth. Table \ref{SW} summarizes the six cases along with the associated atmospheric ion escape rates. We select two locations in the orbit, namely, P$_\mathrm{min}$ and P$_\mathrm{max}$, because they ought to yield the minimum and maximum ion escape rates \citep[e.g.][]{lammer09,DLMC,PFR20}, thus providing the range of values associated with this system. Furthermore, we hold the orientation of the planetary dipole fixed with respect to the IMF, as varying the orientation causes the escape rates to change only by a modest factor of $\lesssim 2$ \citep{DHL19}. As with previous publications, a fiducial 1 bar atmosphere is chosen for both Venus- and Earth-like cases \citep[e.g.][]{DHL17,DJL18}.

\begin{table*}
\caption{Stellar wind input parameters and the associated atmospheric ion escape rates at TOI-700 d for P$_\mathrm{min}$ and P$_\mathrm{max}$, which respectively correspond to the minimum and maximum total stellar wind pressure (P$_\mathrm{tot}$) over one orbital period of TOI-700 d. The photoionization frequencies are rescaled to TOI-700 d values, using the X-ray luminosity from \citet{GBS20} and the empirical EUV scaling from \citet{SMR11}. The planetary mass (1.72 M$_{\oplus}$), radius (1.19 R$_{\oplus}$), orbital period (37.426 days), and semimajor axis (0.163 AU) are adopted from \citet{GBS20}.} \label{SW}
\centering
\begin{tabular}{lllllllll}
\hline
\hline
 Case \# & $P_{sw}$ & n$_{sw}$ & v$_{sw}$ & IMF  & Planetary & Magnetic & O$^+$ Loss Rate & Total Loss Rate  \\
&  & (cm$^{-3}$) & (km s$^{-1}$) & (nT) & Type & Field & (s$^{-1}$) & (s$^{-1}$) \\
\hline
Case 1 &  &  &   & & Venus-like & Off & 1.93$\times$10$^{26}$ & 2.10$\times$10$^{26}$ \\ 
Case 2 & P$_\mathrm{min}$ & 96.8  & (-650.4, 47.4, -7.4)  & (-46.7, 2.9, -3.7) & Earth-like & Off & 1.18$\times$10$^{28}$ & 1.18$\times$10$^{28}$ \\
Case 3 & &  &   & & Earth-like & On & 1.39$\times$10$^{27}$ & 1.39$\times$10$^{27}$ \\ 
\hline
Case 4 &  &  &   & & Venus-like & Off & 2.04$\times$10$^{26}$ & 2.41$\times$10$^{26}$ \\ 
Case 5 & P$_\mathrm{max}$ & 451.6  &  (-471.2, 47.4, 4.3) & (9.2, 2.7, -7.0) & Earth-like & Off & 1.61$\times$10$^{28}$ & 1.61$\times$10$^{28}$ \\
Case 6 & &  &   & & Earth-like & On & 2.12$\times$10$^{27}$ & 2.12$\times$10$^{27}$ \\ 
\hline
\hline
\end{tabular}
\end{table*}

\section{Results}\label{SecRes}
We describe the chief results from our simulations of the stellar wind and atmospheric escape.

\subsection{Stellar wind and radio emission}
The modeled characteristics of the stellar wind were already presented in Section \ref{SSecSW}. In addition, there is another key stellar characteristic that we numerically compute. Based on our MHD numerical simulations, we estimate a stellar mass-loss rate of $\dot{\mathrm{M}}_\star \approx 1.3 \times 10^{-14}\,\mathrm{M}_\odot\,\mathrm{yr}^{-1}$ for TOI-700. It is instructive to compare this result with the stellar mass-loss rate calculated using the semi-analytical approach prescribed in \citet{JGL15}:
\begin{eqnarray}\label{MLPres}
&& \dot{M}_\star \approx 1.4 \times 10^{-14}\,M_\odot\,\mathrm{yr}^{-1}\left(\frac{\Omega_\star}{\Omega_\odot}\right)^{1.33} \nonumber \\
&& \hspace{0.5in} \times\, \left(\frac{M_\star}{M_\odot}\right)^{-3.36}\left(\frac{R_\star}{R_\odot}\right)^2,
\end{eqnarray}
where $R_\star$, $\Omega_\star$, and $M_\star$ are the radius, rotation rate and mass of the star, respectively. For TOI-700, we adopt the stellar parameters from \citet[Table 1]{GBS20} and substitute them into (\ref{MLPres}), there obtaining $\dot{\mathrm{M}}_\star \approx 1.9 \times 10^{-14}\,\mathrm{M}_\odot\,\mathrm{yr}^{-1}$. This estimate exhibits good agreement with the prior numerical prediction for $\dot{\mathrm{M}}_\star$ because the two results differ by a factor of $< 1.5$. 

The calculation of the stellar wind parameters for TOI-700 enable us to address another crucial issue: the detection of the putative magnetic field of TOI-700 d via radio emission generated by the cyclotron maser instability \citep{Zark07}. The frequency $\nu$ at which maximal emission occurs is
\begin{equation}
\nu \approx \frac{e B_p}{2\pi m_e c} \approx  2.8\,\mathrm{MHz}\,\left(\frac{B_p}{1\,\mathrm{G}}\right), 
\end{equation}
where $B_p$ is the planetary magnetic field, while $e$ and $m_e$ denote the electron charge and mass, respectively. As the expected frequency is $< 10$ MHz, an ultra-low frequency radio telescope in space or the lunar surface is necessary to detect such emission.

To determine whether the radio emission is detectable, the radio flux density ($S$) is estimated as
\begin{equation} \label{RFD}
    S = \frac{P_R}{4\pi \Delta{\nu} d^2},
\end{equation}
where $d$ is the Earth-to-planet distance, $P_R$ signifies the planetary radio power, and $\Delta{\nu} \approx \nu/2$ \citep{Zark07}. As the radio emission is potentially dictated by the stellar wind's magnetic power incident on the planet, we employ \citet{VD17} to determine $P_R$:
\begin{equation} \label{RadP}
    P_R \sim 2 \times 10^{-3}\,\left(\frac{\pi \mathrm{B}_{sw}^2 R_M^2 \mathrm{v}_{sw}}{4\pi}\right),
\end{equation}
where $\mathrm{B}_{sw}$ denotes the interplanetary magnetic field (IMF), $\mathrm{v}_{sw}$ is the velocity of the stellar wind, and $R_M$ represents the magnetospheric radius given by
\begin{equation} \label{Mag}
    R_M = R_p \left(\frac{B_p^2}{4\pi \mathrm{P}_{sw}}\right)^{1/6},
\end{equation}
where $\mathrm{P}_{sw} \approx m_p \mathrm{n}_{sw} \mathrm{v}_{sw}^2$ is the dynamical pressure, with $\mathrm{n}_{sw}$ representing the stellar wind density and $m_p$ is the proton mass. After making use of (\ref{RFD})-(\ref{Mag}) along with Table \ref{SW} and R$_p \approx$ 1.2 R$_\oplus$, we are in a position to calculate $S$.  For the maximum pressure (P$_\mathrm{max}$) scenario in Table \ref{SW}, after simplification we have
\begin{equation}\label{Sfin}
    S \approx 1.4 \times 10^{-7}\,\mathrm{Jy}\,\left(\frac{B_p}{1\,\mathrm{G}}\right)^{-1/3}.
\end{equation}
While the estimate is smaller than the predicted radio emission of some nearby exoplanets by orders of magnitude, two essential factors must be noted. As seen from (\ref{Sfin}), the radio emission could increase by an order of magnitude if $B_p$ is a few mG. More importantly, during large CMEs \citep[e.g.,][]{jakosky15,Luhmann2017}, $S$ would increase by a couple of orders of magnitude \citep{DJL18}, thereby potentially attaining values as high as $\sim 0.1$ mJy for weak magnetic fields.

\begin{figure*}
\begin{center}
\includegraphics[scale=0.8]{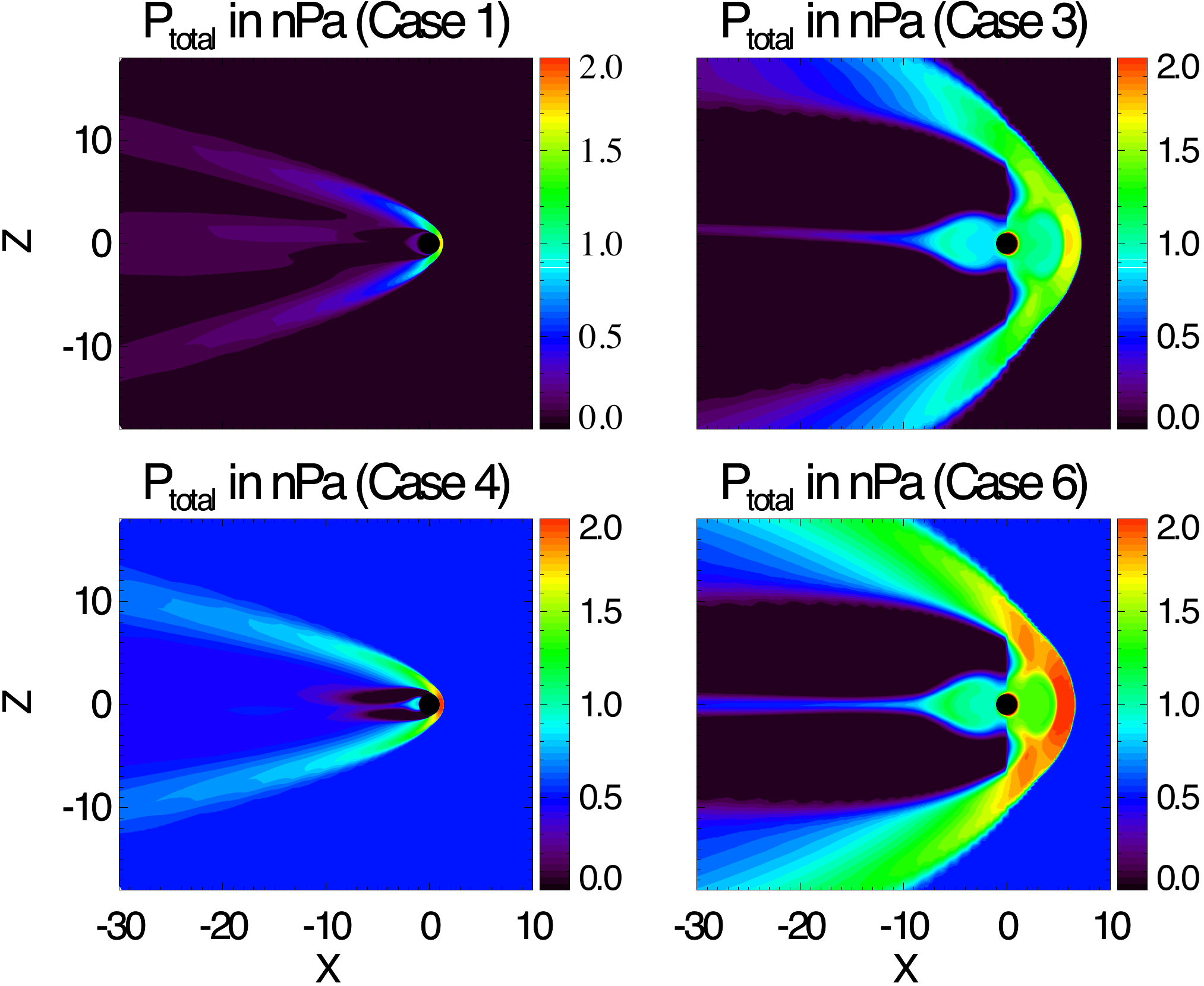}
\end{center}
\caption{Logarithmic scale contour plots of the total pressure (P$_{total}$, units of nPa) - the sum of the ion (P$_{ion}$) and electron (P$_{electron}$) pressure - in the meridional plane based on the stellar wind conditions at P$_\mathrm{min}$ (first row) and P$_\mathrm{max}$ (second row). The first column shows the unmagnetized Venus-like case, while the second column depicts the magnetized Earth-like case.}
\label{fig:pressure}
\end{figure*}

\begin{figure*}
\begin{center}
\includegraphics[scale=0.85]{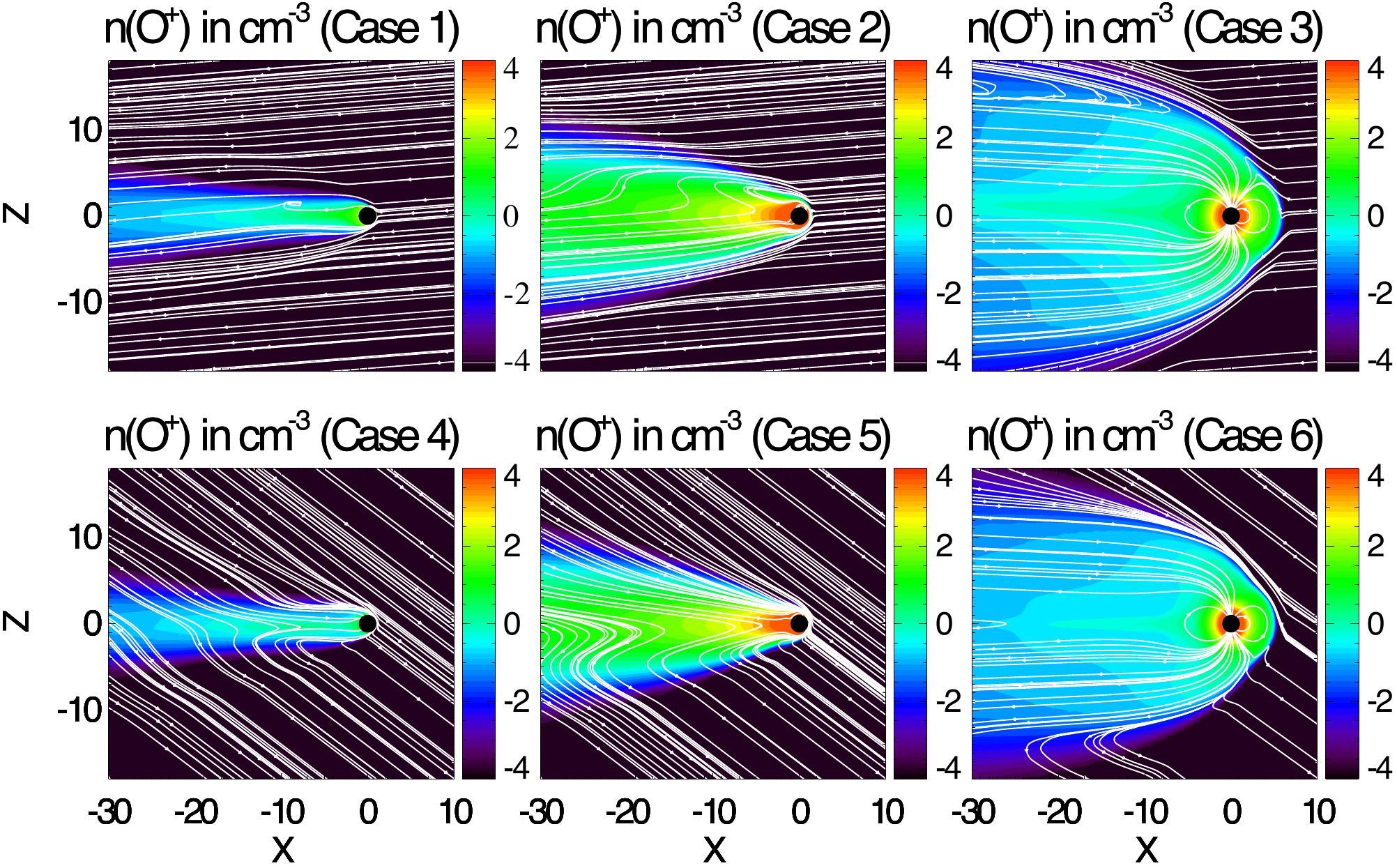}
\end{center}
\caption{Logarithmic scale contour plots of the O$^+$ ion density (units of cm$^{-3}$) with magnetic field lines (in white) in the meridional plane based on the stellar wind conditions at P$_\mathrm{min}$ (first row) and P$_\mathrm{max}$ (second row). The first column shows the unmagnetized Venus-like cases, whereas the second and third columns depict the unmagnetized and magnetized Earth-like cases.}
\label{fig:atmosescape}
\end{figure*}

\subsection{Atmospheric escape}
Next, we direct our attention to the central issue of atmospheric ion escape facilitated by the stellar wind in conjunction with the stellar radiation. 

Figure \ref{fig:pressure} depicts the pressure balance arising from the stellar wind-planet interactions. The first and second columns present the unmagnetized Venus-like and magnetized Earth-like cases, respectively. As a result of the supermagnetosonic nature of the stellar wind at TOI-700 d, bow shocks are formed ahead of this planet, regardless of the inclusion of the planetary magnetic field. Compared to the unmagnetized Venus analog, the presence of the global magnetic field results in a larger interaction cross-section with the stellar wind for the magnetized Earth-like case. An inspection of the bottom panels of Figure \ref{fig:pressure} reveals that the planetary magnetosphere (or induced magnetosphere for unmagnetized cases) is compressed, and the total pressure becomes intensified due to the stronger stellar wind pressure. 

The salient results concerning atmospheric escape are depicted in Figure \ref{fig:atmosescape}, which shows the calculated oxygen ion density with the associated magnetic field lines in the meridional plane for all six cases. This figure yields some general conclusions. To begin with, as seen from Table \ref{SW}, we note that O$^+$ constitutes the dominant ion species undergoing escape for all configurations considered here. The atmospheric oxygen escape rates vary from $\calo(10^{26})$ s$^{-1}$ to $\calo(10^{28})$ s$^{-1}$, indicating that they are higher by a few orders of magnitude than the typical escape rates of $\calo(10^{25}$) s$^{-1}$ for the terrestrial planets in our Solar system \citep[e.g][]{DLMC,DLY18}.

Let us first compare the unmagnetized cases, namely, the first and second columns. We find that the Earth-like case exhibits a stronger and broader flux of escaping O$^+$ compared to the Venus-like case; this result is also consistent with the atmospheric ion escape rates shown in Table \ref{SW}. The reason chiefly stems from the fact that the upper atmosphere of a Venus analog is cooler than its Earth-like counterpart due to the efficient CO$_2$ cooling caused by $15$ $\mu$m emission \citep{bougher08}. In consequence, the exobase of a Venus-like planet is situated lower than that of an Earth-analog, thereby making the former more tightly confined. In other words, the extent of the atmospheric reservoir that is susceptible to erosion by stellar wind is smaller for the Venus analog as compared to an Earth-like atmosphere. 

Now, let us hold the atmospheric composition fixed and vary the magnetic field, i.e., we compare the second column (unmagnetized Earth-analog) with the third column (magnetized Earth-analog). Despite the fact that the magnetized cases exhibit a larger interaction cross-section with the stellar wind (also refer to Figure \ref{fig:pressure}), the planetary magnetic field exerts a net shielding effect for the configurations studied herein for TOI-700 d. The presence of the global magnetic field reduces the atmospheric loss rate by roughly one order of magnitude relative to the unmagnetized case (see Table \ref{SW}). Hence, of the three different scenarios considered in this work, unmagnetized Earth-like worlds are characterized by the highest atmospheric ion escape rates.

\section{Conclusion}\label{SecConc}
In this Letter, we have analyzed the stellar wind parameters and rates of atmospheric escape associated with TOI-700 d, an Earth-sized planet in the HZ of an early-type M-dwarf. In the process, we also calculated the radio emission that would arise from this planet if it possessed an intrinsic dipole magnetic field. Hence, detecting this emission via low-frequency radio observations would enable an estimation of the field strength.

In the case of an unmagnetized Earth-like planet, our simulations imply that a $1$ bar atmosphere may undergo complete depletion in $< 1$ Gyr. In comparison, the available observational constraints on the age of the host star indicate that its age is $> 1.5$ Gyr \citep{GBS20}. On the other hand, when a magnetic field is included and the composition is held fixed, the corresponding time for depletion increases to a few Gyr. Lastly, if an Earth-like atmosphere is replaced with a Venus analog, we estimate that the timescale over which a $1$ bar atmosphere is lost will be $> 10$ Gyr; in fact, the duration exceeds the current age of the Universe. Note, however, that the above estimates are only applicable in the absence of other atmospheric sources and sinks. 

Hence, depending on the composition and magnetic field, TOI-700 d may possess a sizable atmosphere or be devoid of it. By measuring the peak-to-trough amplitude of thermal phase curves, it is possible in principle to discern whether this world has an atmosphere \citep{SWF11}. This method was recently employed by \citet{KKM19} to deduce that the M-dwarf exoplanet LHS 3844b is probably airless, or has a very tenuous atmosphere. However, this conclusion was facilitated by the large peak-to-trough amplitude (of order $100$ ppm); the host star LHS 3844 was also situated closer to Earth. In contrast, if the expected amplitude is an order of magnitude smaller, akin to what might be valid for Proxima b \citep{KL16}, it seems plausible that thermal phase curve measurements by the James Webb Space Telescope are rendered impractical \citep{SWK20}; this may, perhaps, fall under the purview of next-generation telescopes.
 
A few notable caveats are worth pointing out here. For starters, the escape rates were probably higher when the star was younger due to intense stellar winds and EUV fluxes \citep[e.g.][]{lillis15,DLY18}. Hence, at first glimpse, the preceding estimates for atmospheric retention seem to constitute upper bounds. On the other hand, there are several age-dependent geological and putative biological processes that were not accounted for, including outgassing of reducing gases, continental weathering, and the burial of pyrite \citep[e.g.][]{CK17}. Another crucial abiotic process whose significance remains unclear is the extent of H$_2$O photolysis \citep[e.g.][]{TGJ18}, driven by the higher incident X-ray and UV fluxes at TOI-700 d \citep{GBS20}, that might produce a thick oxygen atmosphere for this specific planet, as pointed out by \citet{Lin20}.

In summary, our work underscored a couple of key findings. The subtle role of factors like atmospheric composition and magnetic field in regulating atmospheric escape rates was explicated. Furthermore, we showed that retaining a $1$ bar atmosphere over Gyr timescales is potentially feasible for TOI-700 d under certain circumstances. Hence, this improves the prospects for long-term habitability of this world, and possibly for similar exoplanets around quiescent early M-dwarfs.



\acknowledgments
CD and MJ were supported by NASA grant 80NSSC18K0288. Resources for this work were provided by the NASA High-End Computing (HEC) Program through the NASA Advanced Supercomputing (NAS) Division at Ames Research Center. The Space Weather Modeling Framework that contains the BATS-R-US code used in this study is publicly available from \url{http://csem.engin.umich.edu/tools/swmf}. For distribution of model results in this study, please contact the corresponding author.


\end{document}